\documentclass{article}

\usepackage[utf8]{inputenc}
\usepackage[T1]{fontenc}
\usepackage{lmodern}
\usepackage{verbatim}

\usepackage{amsmath}
\usepackage{hyperref}
\usepackage{subfig}
\usepackage{caption}

\usepackage{epsf}
\usepackage{complex-systems}
\usepackage{graphicx}

\graphicspath{{images/}}
\usepackage[rightcaption]{sidecap}
\usepackage{wrapfig}
\usepackage{subfig}
\usepackage{seqsplit}
\begin{document}

\title{Search of Complex Binary
Cellular Automata Using Behavioral Metrics%
% Use \thanks for footnotes to the title. Put acknowledgments at the
% end of the paper.
}

\author{\authname{Juan C. López-González}
\\[2pt]
% Use \\[2pt] to end the line and add space between author and affiliation
\authadd{Wolfram Research}\\
\authadd{and}\\
\authadd{Physics and Mathematics in Biomedicine Consortium, UCV}\\
\authadd{jlopez@cellular-automata.com}\\
%\authadd{and}\\
%\authadd{Group, Company, Address, City, State ZIP/Zone, Country}\\
\and
% precede the second set of authors with \and.
\authname{Antonio Rueda-Toicen} \\
% Each author name goes on a separate line.
\authadd{Instituto Nacional de Bioingeniería}\\
\authadd{Universidad Central de Venezuela}\\
\authadd{and}\\
\authadd{Physics and Mathematics in Biomedicine Consortium, UCV}\\
\authadd{antonio.rueda@ciens.ucv.ve}
% Put no ``.'' at the end of any address.
}

\maketitle
% End title section

% The following specifies the running headings
\markboth{Juan López-González, Antonio Rueda-Toicen}
{Search of Complex CA Using Behavioral Metrics}

% Each running heading should be less than about 50 characters long.
% If necessary, give a shortened version of the title. Do not use
% abbreviations.
%
% If they will fit, include complete author names; otherwise use
% initials only. If abbreviated names do not fit, truncate the author
% list and end with ``and others''.

\begin{abstract}
We propose the characterization of binary cellular automata using a set of behavioral metrics that are applied to the minimal Boolean form of a cellular automaton's transition function. These behavioral metrics are formulated to satisfy heuristic criteria derived from elementary cellular automata. Behaviors characterized through these metrics are growth, decrease, chaoticity, and stability. From these metrics, two measures of global behavior are calculated: 1) a static measure that considers all possible input patterns and counts the occurrence of the proposed metrics in the truth table of the minimal Boolean form of the automaton; 2) a dynamic measure, corresponding to the mean of the behavioral metrics in \textit{n} executions of the automaton, starting from \textit{n} random initial states. We use these measures to characterize a cellular automaton and guide a genetic search algorithm, which selects cellular automata similar to the Game of Life. Using this method, we found an extensive set of complex binary cellular automata with interesting properties, including self-replication.
\end{abstract}

\section{Introduction}
\label{intro}% A tag for referring to this section (optional).
Cellular automata with complex behavior exhibit dynamical patterns that can be interpreted as the movement of particles through a physical medium. These particles are interpretable as loci for information storage, and their movement through space is interpretable as information transfer. The collisions of these particles in the cellular automaton’s lattice are sites of information processing \cite{nks, stat-mech-ca, universality-ca, collision-computing}. Cellular automata with complex behavior have immense potential to describe physical systems and their study has had impact in the design of self-assembling structures \cite{mol-assembly-ca, sierpinski-dna, amorphous-comp, dna-self} and the modelling of biological processes like signaling, division, apoptosis, necrosis and differentiation \cite{rule-simulation, bio-modelling, handbook, biochem-ca, ca-signal}. John Conway’s Game of Life \cite{game-of-life} is the most renowned complex binary cellular automaton, and the archetype used to guide the search methodology for other complex binary cellular automata that we describe in this work.
Previously, complex behavior in binary cellular automata has been characterized through measures such as entropy \cite{universality-ca}, Lyapunov exponents \cite{lyapunov-1, lyapunov-2}, and Kolmogorov-Chaitin complexity \cite{kolmogorov}. We propose the characterization of the behavior of $n$-dimensional cellular automata through heuristic measures derived from the evaluation of their minimal Boolean forms.  This proposed characterization is derived from heuristic criteria validated in elementary cellular automata with simple Boolean forms.  Table \ref{table:ca-boolean-behavior} illustrates the rationale for this characterization showing elementary cellular automata whose Boolean forms are minimally simple, and whose behavior can be unequivocally identified. Cellular behaviors of growth, decrease, and chaoticity are characterized by the Boolean operations \textit{OR}, \textit{AND}, and \textit{XOR}, respectively. The cellular behavior of stability can be characterized by the absence of a Boolean operator or the use of the \textit{NOT} operator.

\begin{table}
    \resizebox{\textwidth}{!}{%
	\centering
	\begin{tabular}{|c|c|c|c|}
	\hline
	Rule & Sample Evolution & Boolean Form & Behavior\\
	\hline
	$R_{204}$ & \raisebox{-.5\height}{\includegraphics{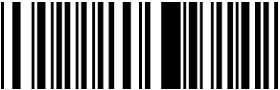}} & \textit{q} & Stable\\
	\hline
	$R_{160}$  & \raisebox{-.5\height}{\includegraphics{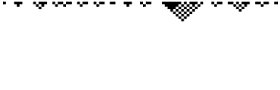}} &\textit{p AND r} & Decreasing\\
	\hline
	$R_{252}$  & \raisebox{-.5\height}{\includegraphics{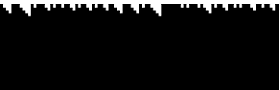}} & \textit{p OR q} & Growing\\
	\hline
	$R_{90}$  & \raisebox{-.5\height}{\includegraphics{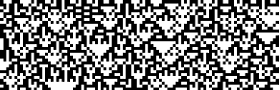}} & \textit{p XOR q} & Chaotic\\
	\hline
	\end{tabular}}
	\caption{Elementary cellular automata with simple Boolean forms, which are unequivocally associated to a particular behavior. The Boolean values of the cells in the neighborhood are $p$ for the left neighbor, $q$ for the central cell, and $r$ for the right neighbor. Black cells are in 1 state, white cells are in 0 state.}
\label{table:ca-boolean-behavior}
\end{table}

We define an evaluation criterion to produce metrics that characterize the behavior of cellular automata whose minimal Boolean expressions are more complex (i.e. have more terms and the combination of various operators) than those appearing in Table \ref{table:ca-boolean-behavior}. The produced metrics are used to create static and dynamic measures of behavior. The static measure of behavior is calculated from the truth table of the minimal Boolean expression of the cellular automaton, and the dynamic measure of behavior is derived from the averaged appearance of the metrics in \textit{n} executions of the cellular automaton from \textit{n} random initial conditions. We use the Euclidean distance of these measures in a given cellular automaton to the measures of the Game of Life to assess its capacity for complex behavior, and use this distance as a cost function to guide the genetic search of $n$-dimensional cellular automata with complex behavior. 

%There may be subsections as well as sections.
%The subsections may, but need not, be numbered.
% \subsection{title} is used for a numbered subsection;
% \subsection*{title} is used for an unnumbered one.
%There should not be subsubsections. Only the first word and proper
%names should be capitalized in section titles.

\section{Definition of binary cellular automaton}
A cellular automaton is formally represented by a quadruple $\{Z,S,N,f\}$, where

\begin{itemize}
\item$Z$ is the finite or infinite cell lattice,
\item$S$ is a finite set of states or values for the cells,
\item$N$ is the finite cell neighborhood,
\item$f$ is the local transition function, defined by the state transition rule.
\end{itemize}

Each cell in the lattice $Z$ is defined by its discrete position (an integer number for each dimension) and by its discrete state value $S$. In a binary cellular automaton,  $S = \{0,1\}$.
Time is also discrete. The state of the cell is determined by the evaluation of the local transition function on the cell’s neighborhood at time $t$;  $t + 1$ is the next time step after time $t$. The neighborhood is defined as a finite group of cells surrounding and/or including the observed cell.

\subsection{Lattice, cell and configuration}
The global state is the configuration of all the cells that comprise the automaton, $C \in S^Z$. The lattice $Z$ is the infinite cyclic group of integers $\{ \dots ,-1,0,1,2, \dots \}$. The position of each cell in the lattice is described by the index position $x \in Z$. Configurations are commonly written as sequences of characters, such as

\begin{equation}
C = \dots c_{-1} c_0 c_1 c_2 \dots
\end{equation}

The finite global state is a finite configuration $C \in S^Z$, where $Z$ is a finite lattice, indexed with ${0,1,2,3 \dots n-1}$ integers,

\begin{equation}
C = c_1 c_2 \dots c_x c_{x+1} \dots c_{n-2} c_{n-1}
\end{equation}

\subsection{Neighborhood and local transition function}
The set of neighborhood indices $A$ of size $m = |A|$
is defined by the set of relative positions within the configuration, such that

\begin{equation}
A={a_0,a_1, \dots ,a_{m-2},a_{m-1}}
\end{equation}

$N_x$ is the neighborhood of the observed cell $c_x$ that includes the set $A$ of indices, and is defined as

\begin{equation}
N_x = c_{x+a_0} c_{x+a_1} \dots c_{x+a_{m-2}} c_x + a_{m-1}
\end{equation}

this describes the neighborhood as a character string that includes the cells
that are considered neighbors of the observed cell $x$. A compact representation of the neighborhood value $N_x$ is a unique integer, defined as an $m-$digits, $k-$based number [2]

\begin{equation}
N_x = \sum_{i=0}^{m-1}k^{m-1-i}c_{x+a_i} = c_{x+a_0}k^{m-1} + \dots +c_{x+a_{m-1}k}k^0
\end{equation}

The local transition function $f$ yields the value of $c_x$ at $t+1$ from
the neighborhood of the cell observed at present time $t$ is expressed by

\begin{equation}
f(N_x^t )= c_x^{t+1}
\end{equation}

where $N_x^t$ specifies the states of the neighboring cells to the cell $x$ at time $t$. The transition table defines the local transition function, listing an output value for each input configuration. Table \ref{table:tran-function-truth-table} is a sample transition table for an elementary cellular automaton with a neighborhood of radius 1, wherein adjacent neighboring cells of  $c_x$   are   $c_{x-1}$ and $c_{x+1}$, forming a tuple $\{c_{x-1},c_x,c_{x+1}\}$, $S\in\{0,1\}$.

\begin{table}
		\centering
	\begin{tabular}{|c|c|} 
		\hline
		$N_x^t$ & $f(N_x^t)$  \\ 
		\hline
		000 & 0  \\
		\hline 
		001 & 1  \\ 
		\hline
		010 & 1 \\
		\hline
		011 & 1 \\
		\hline
		100 & 1 \\
		\hline
		101 & 0 \\
		\hline
		110 & 1 \\
		\hline
		111 & 0 \\
		\hline
	\end{tabular}
	\caption{Local transition function of $R_{94}$ as a truth table.}
	\label{table:tran-function-truth-table}
	\end{table}

\subsection{Global transition function}
\label{global-transition-function}
The global dynamics of the cellular automaton are described by the global transition function $F$

\begin{equation}
F:S^N \rightarrow S^N
\end{equation}

$F$ is the transition between the current global configuration $C^t$ and the next global configuration $C^{t+1}$

\begin{equation}
C^{t+1}=F(C^t)
\end{equation}
 	
The global transition function $F$ is defined by the local transition function $f$ as

\begin{equation}
F(C_x)= \dots f(N_{x-1} )f(N_x )f(N_{x+1} ) \dots 	
\end{equation}

\section{Transformation of the cellular space}
We redefine the local transition function to incorporate behavioral knowledge of the automaton’s evolution, given an input/output pair. This redefined function is applied to all cells of the automaton at a given evolution step t to quantify its overall behavior.

\subsection{Redefined local transition function}
The redefined local transition function $g$ calculates the behavioral metric of a single cell  $c_x$ evaluating the local transition function $f$ on its neighborhood $N_x^t$. Through the local transition function $g$, we define the transformation $d_x^{t+1}$ that yields the next step of the evolution of cell $c_x$ as

\begin{equation}
d_x^{t+1}=g(f,N_x^t )
\end{equation}

This transformation is necessary to calculate the measure of dynamic behavior during the automaton’s evolution, and we propose the inclusion of a metric characterizing the cell behavior obtained after evaluating a particular input.
Input for the Boolean operators considered may be of the form

\begin{equation}
\mathit{Input_1   <operator> Input_2 = Output} 		
\end{equation}

where
$\mathit{<operator>} \, \in \{\mathit{OR,AND,XOR}\}$
The behaviors associated with each binary Boolean operator and its possible inputs and outputs are shown in Table \ref{table:binary-behavior}.

\begin{table}[h!]
		\centering
		\begin{tabular}{ |c|c|c|c|c|c|c|} 
			\hline
			$\mathit{Input_1}$ & $\mathit{Input_2}$ & $\mathit{Output}$ & Behavior & $\mathit{OR}$ & $\mathit{AND}$ & $\mathit{XOR}$\\ 
			\hline
			0 & 0 & 0 & Stability & X & X & \\
			\hline 
			1 & 0 & 0 & Decrease &  & X & \\
			\hline
		    0 & 1 & 0 & Decrease &  & X & \\
			\hline
		    1 & 1 & 0 & Chaoticity &  &  & X \\
			\hline
			 0 & 0 & 1 & Chaoticity &  &  & X \\
			\hline
			 1 & 0 & 1 & Growth & X &  &  \\ 
			 \hline
			 0 & 1 & 1 & Growth &  &  & X  \\
			 \hline 
			 1 & 1 & 1 & Stability & X & X &   \\
			 \hline
		\end{tabular}
		\caption{Behaviors associated to binary Boolean patterns}
		\label{table:binary-behavior}
	\end{table}

The behaviors associated with unary patterns are shown in Table \ref{table:unary-behavior}.

\begin{equation}
\mathit{<operator> Input = Output}
\end{equation}

where $\mathit{<operator>} \, \in \{\mathit{NOT, NOP}\}$

\begin{table}[h!]
		\centering
		\begin{tabular}{ |c|c|c|c|c|} 
			\hline
			$\mathit{Input}$ & $\mathit{Output}$ & Behavior & $\mathit{NOT}$ & $\mathit{NOP}$\\ 
			\hline
			1 & 1 & Stability &  & X \\
			\hline 
			0 & 0 & Stability &  & X \\
			\hline
		    1 & 0 & Stability & X &  \\
			\hline
		    0 & 1 & Stability & X &  \\
			\hline
		\end{tabular}
		\caption{Behavior associated to unary Boolean patterns}
		\label{table:unary-behavior}
	\end{table}
	
where \textit{NOP} stands for “no operator”.
To characterize the automaton’s behavior, we expand the state space

\begin{equation}
g:\{S^N,f\}\rightarrow M,
\end{equation} 	 		

  where

\begin{equation}
M = \{0,1,2,3,4,5\}
\end{equation}

The different values of $M$ abbreviate the duples of state and behavior shown in Table \ref{table:m-code}.  Each tuple is obtained from the result of the local transition function $g$ applied to a particular configuration of the cell $x$ and its neighborhood $N$.

\begin{table}[h!]
		\centering
		\begin{tabular}{ |c|c|} 
			\hline
			$M$ & ${\{S_x^{t+1}, behavior\}}$\\ 
			\hline
			0 & $\{0, \mathit{stable}\}$  \\
			\hline 
			1 & $\{0, \mathit{decreasing}\}$ \\
			\hline
		    2 & $\{0, \mathit{chaotic}\}$  \\
			\hline
		    3 & $\{1, \mathit{chaotic}\}$  \\
			\hline
			4 & $\{1, \mathit{growing}\}$ \\
			\hline
			5 & $\{1, \mathit{stable}\}$\\
			\hline
		\end{tabular}
		\caption{$M$ code, abbreviation of duples of cell state and behavior obtained when applying the local transition function $g$.}
		\label{table:m-code}
	\end{table}

The $M$ code eases the implementation of an algorithmic search for cellular automata with interesting behavior using the proposed metrics. According to the $M$ code, chaotic and stable behaviors may generate 1 or 0 as output from 1 or 0 as input, growing behavior may only generate 1 as output from 0 as input, and decreasing behavior may only generate 0 as output from 1 as input.

\subsection{Global transition function}

The global behavioral metric of the cellular automaton is characterized as 

\begin{equation}
G:\{S^N,f \} \rightarrow M^N
\end{equation}

$G$ represents the transition between the current global configuration  $C^t$
and the next global configuration $C^{t+1}$. We set $D^0 = (0,f)$ and express the 
automaton's global behavioral metric as 

\begin{equation}
D^{t+1} = G(C^t,f)
\end{equation} 

for example, from the initial state, 

$$C^0(initial\,state)$$
$$C^1 = F(C^0) \rightarrow D^1 = G(C^0, f)$$
$$C^2 = F(C^1) \rightarrow D^2 = G(C^1, f)$$
$$C^3 = F(C^2) \rightarrow D^3 = G(C^2, f)$$
$$C^4 = F(C^3) \rightarrow D^4 = G(C^3, f)$$ 
$$\vdots$$

The redefined global transition function $G$ is expressed as the concatenated string obtained when the redefined local transition function $g$ is applied to all of the automaton’s cells $c_i$   

\begin{equation}
G(\dots c_{x-1} c_x c_{x+1} , f) = \ldots g(n_{x-1} , f) g(n_x , f) g(n_{x+1}, f) \dots
\end{equation}

\subsection{Implementation of $g(f, N_x^t)$}

The $g$ function incorporates heuristic information that enables the measurement of behaviors in the automaton’s lattice. 
The $g$ function performs the following steps, given a pattern $N_x^t$  and the transition function $f$:

\begin{enumerate}
\item The local transition function $f$ is simplified to its minimal Boolean expression. 
\item $f$ is expressed as a binary execution tree.
\item $N_x^t$  is evaluated on the binary execution tree obtained in $2$.
\end{enumerate}

In Table \ref{table:ca-boolean-behavior} we mentioned the behavioral characterization corresponding to cellular automata whose minimal expression correspond to a single Boolean operator. This characterization needs to be extended to describe cellular automata whose minimal forms have several distinct Boolean operators. To tackle this problem, we express a cellular automaton’s transition function in a binary evaluation tree and propose a set of evaluation rules for its nodes based on heuristic criteria.

We write the transition function of the minimal expression of the automaton’s rule in a tree graph. We assign to each node of the tree a Boolean operation. The transition function is evaluated, with input placed at the tree's leaves, according to heuristic rules. The result of the evaluation is obtained at the root node. The heuristics considered are crafted to fit criteria derived from the characteristic behaviors of several elementary cellular automata.  

The proposed heuristic $H$ consists of rules for evaluation of the nodes in the binary tree. These tree evaluation rules are defined for 
$${\mathit{term} \, \mathit{<OPERATOR>}  \, \mathit{term}}$$      and
$$\mathit{\mathit{<OPERATOR>} \, \mathit{term}}$$

where $\mathit{<OPERATOR> \, \in \, \{ \mathit{AND, OR, XOR, NOT} \}}$ and $\mathit{term}$ corresponds to the set 
$M=\{ 0,1,2,3,4,5 \}$

\begin{figure}[h]
\centering
\begin{tabular}{ l l }
 \includegraphics[scale=0.8]{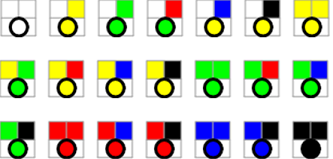} & \includegraphics[scale=0.8]{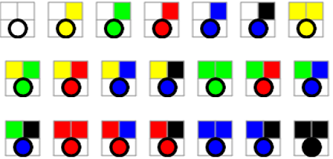}  \\ 
 \textit{AND} & \textit{OR}\\
 \includegraphics[scale=0.8]{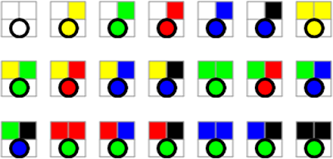} & \includegraphics[scale=0.7]{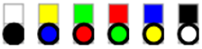}  \\  
 \textit{XOR} & \textit{NOT}\\
\end{tabular}

\caption{Tree Evaluation rules $H$, squares correspond to inputs and circles to outputs. White corresponds to $M = 0 = \{0, \mathit{stable}\}$; yellow corresponds to $M = 1 = \{0, \mathit{decrease}\}$; green corresponds to $M = 2 = \{0, \mathit{chaotic}\}$; red corresponds to $\mathit{M = 3 = \{1, \mathit{chaotic}\}}$; blue corresponds to $M = 4 = \{1, \mathit{growth}\}$; black corresponds to $\mathit{M = 5 = \{1, \mathit{stable}\}}$.}
\label{fig:rules}
\end{figure}

\begin{figure}[h]
\centering
\begin{tabular}{ccc}

 \includegraphics[scale=0.25]{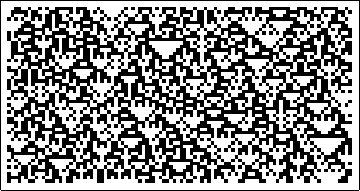}      &
 \includegraphics[scale=0.25]{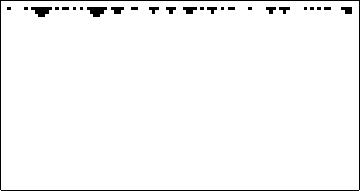}     & 
 \includegraphics[scale=0.25]{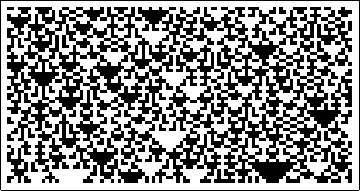}     \\
 \textit{ $R_{90}=$   p XOR r}                     &
 \textit{ $R_{128}=$  p AND q AND r}               &  
 \textit{ $R_{150}=$  p XOR q XOR r}               \\
 
 \includegraphics[scale=0.25]{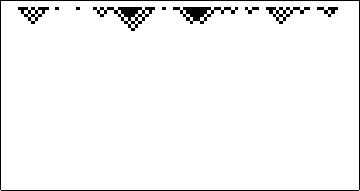}    &
 \includegraphics[scale=0.25]{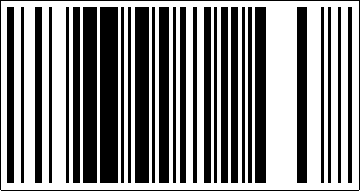}    & 
 \includegraphics[scale=0.25]{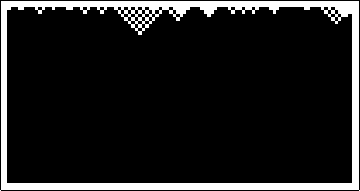}    \\
 \textit{$R_{160}=$ p AND r}                      &
 \textit{$R_{204}=$ q}                            &
 \textit{$R_{250}=$ p OR r}                       \\
                                                   & %blank cell
 \includegraphics[scale=0.25]{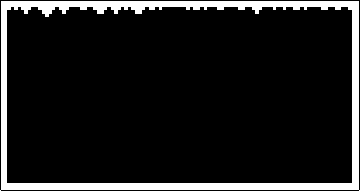}     \\
                                                   & %blank cell
 \textit{$R_{254}=$ p OR q OR r}                   \\
 
\end{tabular}
\caption{Elementary cellular automata used to define the criteria in $H$.}
\label{fig:eca_from_criteria}
\end{figure}

Figure \ref{fig:rules} shows the heuristic precedence rules defined for each logical operator.
Figure \ref{fig:eca_from_criteria} shows the elementary cellular automata used to define the heuristic characterization criteria, alongside their minimal Boolean forms. 
%\begin{comment}
%\end{comment}

\newpage 

\begin{itemize}
\item \textbf{Criterion 1 -} In the leaf nodes, 
$S = 0$ must be equivalent to $M = 0 = \{0, \, \mathit{stable}\}$ and $S = 1$ must be equivalent to $M = 5 = \{1, \mathit{stable}\}$. 
\item \textbf{Criterion 2 -} Chaoticity measured in $R_{150} = p \, \mathit{XOR} \, q \, \mathit{XOR} \, r$ must be greater than chaoticity measured in $R_{90}=p \, \mathit{XOR} \, r$.\\
The proposed heuristic $H$ produces the following behavioral metrics in these automata:\\
$$R_{150} \, \mathit{chaoticity} = 0.375$$
$$R_{90}  \, \mathit{chaoticity} = 0.25$$
\item \textbf{Criterion 3 -} Chaoticity measured in $R_{90}=p \, \mathit{XOR} \, r$ must be greater than chaoticity measured in $R_{204}=q$.\\
The proposed heuristic $H$ produces the following behavioral metrics in these automata:\\
$$ R_{90}  \, \mathit{chaoticity} = 0.25 $$ 
$$ R_{204} \, \mathit{chaoticity} = 0 $$ 
\item \textbf{Criterion 4 -} Decrease measured in $R_{128}=p \, \mathit{AND} \, q \, \mathit{AND} \, r$ must be greater than decrease measured in $R_{160}=p \, \mathit{AND} \, r$.\\
The proposed heuristic $H$ produces the following behavioral metrics in these automata: 
$$ R_{128} \, \mathit{decrease} = 0.75 $$
$$ R_{160} \, \mathit{decrease} = 0.5 $$

\newpage

\item \textbf{Criterion 5 -} Decrease measured in 
$ R_{128}=p \, \mathit{AND} \, q \, \mathit{AND} \, r $ must be greater than 
decrease measured in $ R_{160}=p \, \mathit{AND} \, r $.\\
The proposed heuristic $H$ produces the following behavioral metrics in these automata:
$$ R_{160} \, \mathit{decrease} = 0.5 $$
$$ R_{204} \, \mathit{decrease} =  0 $$
\item \textbf{Criterion 6 -} Growth measured in $R_{254}=p \, \mathit{OR} \, q \, \mathit{OR} \, r$
must be greater than growth measured in $R_{250}=p \, \mathit{OR} \, r$\\
The proposed heuristic $H$ produces the following behavioral metrics in these automata:\\
$$ R_{254} \, \mathit{growth} = 0.75 $$
$$ R_{250} \, \mathit{growth} = 0.5 $$
\end{itemize}

\begin{figure}[h!]
\centering
\includegraphics[scale=0.6]{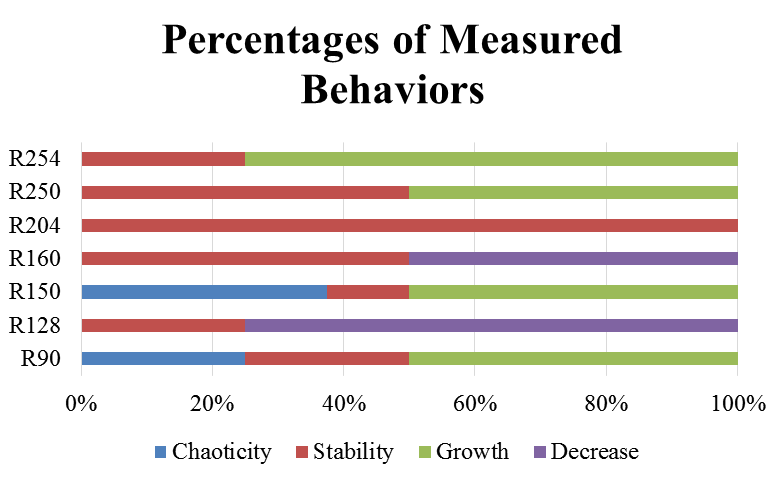}
\caption{Behavioral percentages in elementary cellular automata considered as criteria for evaluating the proposed heuristic}
\label{fig:behavior-percentages}
\end{figure}

Figure \ref{fig:behavior-percentages} shows percentage of measured behaviors, using the proposed set of evaluation rules $H$, in the elementary cellular automata considered in the
criteria.

\newpage

\subsection{Evaluation example with $R_{94}$}

\begin{figure}
\centering
\includegraphics[scale=0.6]{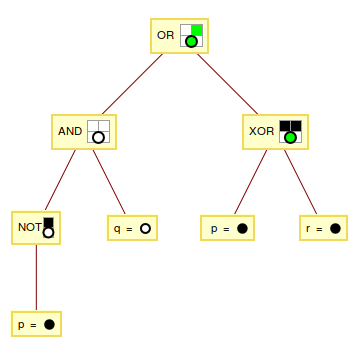}
\caption{Evaluation of the input pattern 101 in $R_{94}$ with the proposed rules.} 
\label{fig:evaluation-tree-example}
\end{figure}

The minimal Boolean expression of $R_{94}$, 
$f = (q \, \mathit{AND} \, (\mathit{NOT} \, p)) \, \mathit{OR} \, \\ (p \, \mathit{XOR} \, r)$, 
is placed  in a binary evaluation tree, as shown in Figure \ref{fig:evaluation-tree-example}. 
Each node in the tree is evaluated using the rules shown in Figure \ref{fig:rules}. 
This process is demonstrated in the Steps 1-5 listed below. 

\begin{comment}
The considered input is the neighborhood 
$N_q^{t=0}= \{ p=1,q= 0,r = 1 \}$. 
\end{comment}

\begin{itemize}

\item \textbf{Step 1.} In the leaf nodes, the values 
$ N_q^{t=0}$ are $\{p=1, q=0, r=1\}$,

are transformed, using the $M$ code mentioned in Section 3.1 as follows:

$$ S_M (p) = \{1, \mathit{stable} \} = 5 $$
$$ S_M (q) = \{0, \mathit{stable} \} = 0 $$
$$ S_M (r) = \{1, \mathit{stable} \} = 5 $$

thus, the input tuple $\{p=1,q=0,r=1\}$ is converted into $\{p=5, \, q=0, \, r=5\}$.

\item \textbf{Step 2.} Leaf $p = 5$ is evaluated at the $\mathit{NOT}$ node, producing
output $ 0 = \{0, \mathit{stable}\} $

\item \textbf{Step 3.} Leaf $q=0$ and the result of step 2 are evaluated at the $\mathit{AND}$ node, producing output $0 = \{0, stable\}$.

\item \textbf{Step 4.} Leaves $p = 5$ and  $r = 5$ are evaluated at the $\mathit{XOR}$ node producing output $2 = \{0, \mathit{chaotic}\}$.

\item \textbf{Step 5.} The output of step 4 and the output of step 5 are evaluated at the $\mathit{OR}$ node, producing as final output $2 = \{0, \mathit{chaotic}\}$. The cell $q$ gets assigned to state 0 in $t+1$, and a counter for the occurrence of chaotic behavior 
in the states of $R_{94}$ would get incremented by one. 
\end{itemize}

\section{\label{section4}Behavioral characterization}

To characterize the overall behavior of a cellular automaton with the proposed metrics, we consider the correlation between two measures: 

\textbf{1)} A static measure, which is the counted occurrence of behaviors associated to the code $M$, in the output of the truth table of the minimal Boolean expression of the cellular automaton.  

\textbf{2)} A dynamic measure, which is the median occurrence of behaviors
associated to the code $M$ in $n$ executions of the cellular automaton, starting from $n$ random initial states. 

\subsection{\label{subsection4.1}Static measure of behavior}

The local transition function transition $f$ is expressed as a truth table, which is converted to $g$ when we include behavioral information. To calculate the static measure of behavior, we count the occurrence of behaviors associated with the values of $M$ in the output of the truth table. This static measure is a vector, with the percentages of chaoticity, stability, growth and decrease measured in the cellular automaton. This static measure is represented as a vector, with the percentages of chaoticity, stability, growth and decrease measured in the cellular automaton. 

For example, in $R_{94}$ the rule is characterized using the $M$ code as shown in Table \ref{table:tt-r94}. 
\begin{table}
		\centering
	\begin{tabular}{|c|c|c|} 
		\hline
		$N_x^t$ & $f(N_x^t)$ & $g(N_x, f)$  \\ 
		\hline
		000 & 0 & $M=1$ \\
		\hline 
		001 & 1 & $M=4$ \\ 
		\hline
		010 & 1 & $M=4$\\
		\hline
		011 & 1 & $M=4$\\
		\hline
		100 & 1 & $M=4$\\
		\hline
		101 & 0 & $M=2$\\
		\hline
		110 & 1 & $M=4$\\
		\hline
		111 & 0 & $M=2$\\
		\hline
	\end{tabular}
	\caption{Truth table of $R_{94}$, with associated $M$ code}
	\label{table:tt-r94}
	\end{table}

To obtain the static measure of $R_{94}$, we count the occurrences of $M$. The static measure of the rule is the percentage of behavioral occurrence in the automaton, as shown in Table \ref{table:static-measure}.

\begin{table}
	\centering
	\begin{tabular}{|c|c|c|c|}
	\hline
	Stability & Decrease & Growth & Chaoticity\\
	\hline
	$M = \{0,5\}$ & $M=1$ & $M=4$ & $M=\{2,3\}$\\
	\hline
	 0\% &  12.5\% &  62.5\% &  25\% \\
	\hline 
	\end{tabular}
	\caption{Behavioral percentages in $R_{94}$, static measure}
	\label{table:static-measure}
\end{table}

We express this measure as a vector of percentages.  

\begin{equation}
M_E = \{ stability \, \% , decrease \, \% , growth \, \% , chaoticity \, \% \}
\end{equation}

For $R_{94}$, the static measure of behavior is

$$M_E = \{0, 12.5, 62.5, 25\}$$

\subsection{\label{subsection 4.2}Dynamic measure of behavior}

To estimate the dynamic measure of behavior $M_D$, we execute the cellular
automaton $n$ times, from $n$ random initial configurations $C_i^{t=0}|i \in n$. We sample occurrences of $M$ in the cell space up to the $k$-th evolution step, where $k$ is an integer > 0, obtained from a uniform distribution.

\begin{equation}
M_D(g) = \lim_{x \to \infty} \frac{(M_D^{t=k}(g, c_i^{t=0}))}{n}
\end{equation}

\begin{figure}
\centering
\includegraphics[scale=0.6]{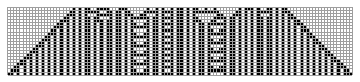}
\includegraphics[scale=0.6]{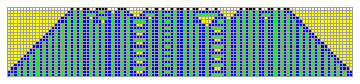}
\caption{Evolution of $R_{94}$ from a random initial configuration, yellow coloring for $M = 1$, green coloring for $M = 2$ and blue coloring for $M = 4$. The code $M$ was applied to cells in  $t \geq 1$. The percentage of cells with $M = 1$ (decreasing behavior) is 18.658\%, cells with $M = 2$ (stable behavior) are 32.467\% and cells with $M = 4$ (chaotic behavior) occupy 48.874\% of the lattice.}
\label{fig:r94-sampling}
\end{figure}

We exclude cells at $t = 0$ from the sampling. The percentages of behavioral occurrences are calculated from the mean of samples. Figure \ref{fig:r94-sampling} shows the sampling of $R_{94}$  in $k = t = 20$. 

\newpage

\section{\label{section 5}Analysis of the Game of Life}

The Game of Life is a complex cellular automaton, class IV according to the classification proposed by Wolfram \cite{nks, universality-ca}. In this cellular automaton, there is a negative correlation between the static measure of behavior and the dynamic measure of behavior. Table \ref{tab:measures} shows this negative correlation and the absolute difference between the static measure and the dynamic measure in the Game of Life.

Some observations pertinent to the measured behavior in the Game of Life: 
\begin{itemize}
\item Static measure: chaotic behavior predominates, an important characteristic of class III automata. 
\item Dynamic measure: decreasing behavior predominates, an important characteristic of class I automata. 
\end{itemize} 

Looking at the transition function  $f$  of The Game of Life, one can find patterns such as

$$(\dots)$$ 
$$ \mathit{NOT} \, x0 \, \mathit{AND} \, \mathit{NOT} \, x1 \, \mathit{AND} \, \mathit{NOT} \, x2 \, \mathit{AND} \, x8 $$ 
$$ \mathit{AND} \,  \, \boldsymbol{(x3 \, \mathit{XOR} \, x4 )} \, \mathit{AND} \,   
\boldsymbol{(x5 \, \mathit{XOR} \, x6)} \, \mathit{OR}$$ 
$$ \, \mathit{NOT} \, x0 \, \mathit{AND} \, \mathit{NOT} \, x1 \, \mathit{AND} \, \mathit{NOT} \, x3 \, \mathit{AND} \, x8 $$ $$ \mathit{AND} \, \boldsymbol{(x2 \, \mathit{XOR} \, x4)} \, \mathit{AND} \, \boldsymbol{(x5 \, \mathit{XOR} \, x6)} \, \mathit{OR}$$ 
$$ (\dots) $$

\begin{table}
	\centering
	\captionsetup{justification=centering}
	\begin{tabular}{|c|c|c|c|}
	\hline
	 & $M_E$ & $M_D$ \\
	\hline
     Chaoticity & 67.96 & 13.38\\
	\hline 
	Decrease & 4.68 & 75.23\\
	\hline
	Growth & 27.34 & 11.37\\
	\hline 
	Stability & 0 & 0\\
	\hline
	\end{tabular}
	\caption{Static and dynamic measures in the Game of Life, their correlation is -0.29}
	\label{tab:GoL-measures}
\end{table}

It is our hypothesis that the emergence of complex behavior in the Game of Life is determined by the appearance of islands of chaotic behavior, surrounded by decreasing patterns. Taking a close look at the boolean expression of $f$ in the Game of Life, one can observe chaotic sub-expressions like $\boldsymbol{(x3 \, \mathit{XOR} \, x4)}$ being "restricted" with \textit{AND-ing} by decreasing sub-expressions such as $( \mathit{AND} \, \mathit{NOT} \, x2 \, \mathit{AND} \, x8)$.

\begin{figure}
\centering
\includegraphics[scale=0.5]{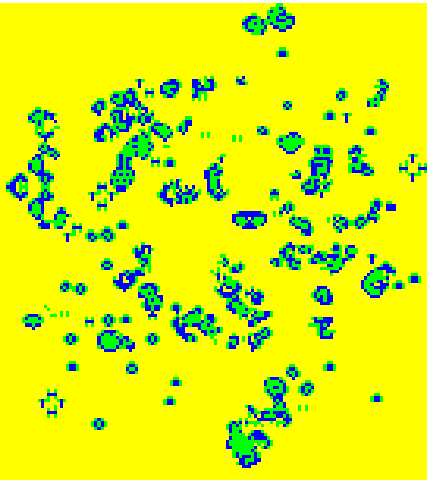}
\caption{The Game of Life, colored according to behavior}
\label{fig:GoL-colored}
\end{figure}
 
In Figure \ref{fig:GoL-colored}, yellow cells have value $M = 1$ (decreasing behavior), and blue cells have value $M=4$ (growth behavior). Green cells cells have $M=2$, exhibiting chaotic behavior. Note that in Figure \ref{fig:GoL-colored}, decreasing cells ($M=1$) cover the largest proportion of the lattice, which corresponds with the dynamic measure of measure of decrease shown in Table \ref{tab:GoL-measures}. One can also appreciate how the isolated patterns exhibit a combination of growth ($M=4$) and chaoticity ($M=2$).

\section{\label{section6} Search of complex binary cellular automata in two dimensions} 
The proposed behavioral metrics were crafted using heuristic criteria from one-dimensional binary cellular automata, yet are applicable to characterize binary cellular automata with different neighborhoods in lattices of higher dimensions. To demonstrate this, we developed a genetic search algorithm \cite{genetic-search} of non-totalistic 2D cellular automata in the Moore neighborhood with radius equal to one. This algorithm searches for automata with behavioral measures similar to those in the Game of Life in a space of size  
$2^{512}$. The genetic algorithm uses a cost function to evaluates each cellular automaton in the population, with this cost being the Euclidean distance between the behavioral measures of each cellular automaton with the behavioral measures of the Game of Life. Another selection condition was added: like the Game of Life, the selected cellular automaton must have $\mathit{stability} = 0$ in both its static and dynamic measures. 
We found a large number of cellular automata with interesting complex behaviors, like gliders, blinkers and self-replicating patterns.

\subsection{\label{section6.1}Tests and results}
\begin{figure}[htbp]
\centering
\begin{tabular}{c}
\subfloat[$t =0$]{\includegraphics[scale=0.55]{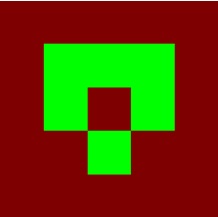}}\\
\subfloat[Replication of the initial state at $t = 91$]{\includegraphics[scale=0.5]{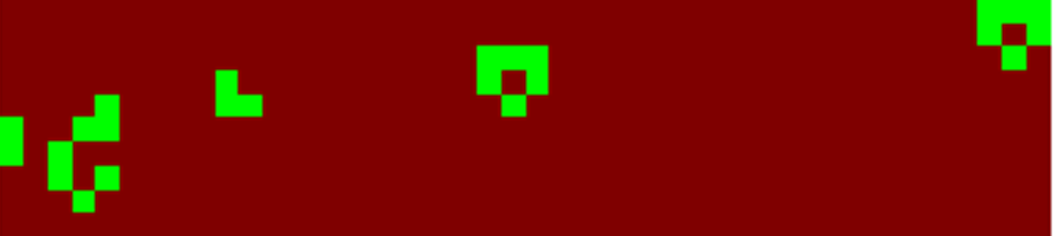}}\\
\subfloat[Persistence of the pattern and its copy at  $t = 307$]{\includegraphics[scale=0.5]{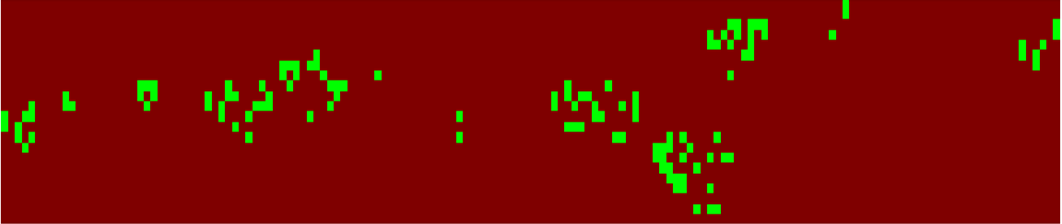}}\\
\end{tabular}
\caption{Self replication in cellular automaton with behavioral metrics similar to the Game of Life, the feature vector of behavioral metrics used to find it is shown below.}
\label{fig:self-replication}
\end{figure}

The proposed genetic search algorithm evolved an initial population of 20 individuals through 5000 generations, each individual being a cellular automaton with a randomly generated transition function $f$. Each cellular automaton's transition function is represented in the population as a chromosome of 512 Boolean values. One point crossover and random mutation (with probability 0.01) were applied at each evolution step\cite{intro-to-ga}. The sampling used to measure dynamic behavior was taken at random intervals at least 10 times for each cellular automaton in the population. In a space of $2^{512}$ possible cellular automata, we generated about 10000 different cellular automata through crossover and mutation, and selected the 1000 closest to the behavioral measures of the Game of Life. These automata were qualitatively evaluated. We found 300 cellular automata in which one can appreciate gliders, blinkers, and other interesting complex behaviors. Among the cellular automata with complex behavior found, we identified a self-replicating cellular automaton, corresponding to Wolfram rule number $168956220003150428540506549680417619769424995409487733442556\\
 339612333081717128579374366701058219674682166161189003344417\\
 08509286446343520818184926824448$.
In this automaton, we can appreciate a pattern that is replicated twice after 91 steps, as shown in Figure \ref{fig:self-replication}.  
Curiously, this complex cellular automaton is more distant from the behavioral measures of the Game of Life than other CA found using the proposed methodology. However, one characteristic is prevalent in this and the other found CA: a negative correlation between their static and dynamic behavioral measures.  
 
 \newpage
 
\begin{table}[h]
	\centering
	%\captionsetup{justification=centering}
	\begin{tabular}{|c|c|c|}
	\hline
	 & $M_E$ & $M_D$ \\
	\hline
	Chaoticity & 61.72 & 5.61  \\
	\hline 
	Decrease   & 3.32 & 90.63  \\
	\hline
	Growth     & 34.96 & 3.77 \\
	\hline 
	Stability  & 0 & 0 \\
	\hline
	\end{tabular}
	\caption*{Euclidean distance to the feature vector of behavioral metrics of the Game of Life is 21.31. The correlation between measures of static and dynamic behavior in this cellular automaton is -0.45}
	
	\label{tab:measures}
\end{table}
 
\newpage 

\begin{figure}[h]
\centering

\centering
\begin{tabular}{ c c}
\includegraphics[scale=0.4]{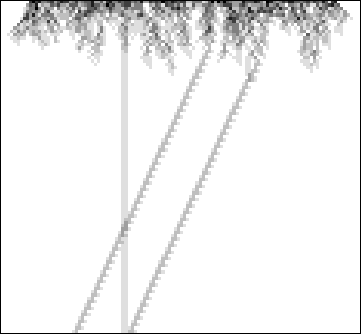} &
\includegraphics[scale=0.4]{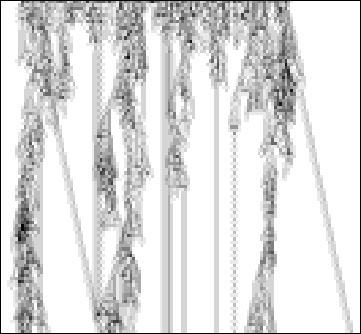}\\
 Found Self-Replicative CA & Game of Life \\
\end{tabular}

\caption{Averaged spacetime evolutions, showing gliders as streaks.}
\label{fig:spacetime-replicative-and-GoL}
\end{figure}

This self-replicative pattern is a particular kind of localized structure that moves across the cellular automaton's lattice. These localized structures, or "gliders" (which aren't always self-replicative) can be seen as streaks in averaged spacetime slices that depict the evolution of the cellular automaton from random initial conditions.  
Figure \ref{fig:spacetime-replicative-and-GoL} shows the spacetime slices depicting gliders on the found self replicative CA and on the Game of Life, as comparison.  

We present examples of complex cellular automata found with the proposed search method\footnotemark. Mean spacetime visualizations of the evolving state of the automaton are provided for each; the lower rows of the lattice being the latter time steps. A list of 277 selected complex binary cellular automata can be found in the Bitbucket repository at \url{http://bit.ly/complexbinaryca}. A Java implementation of the genetic search algorithm based on behavioral metrics is available at \url{http://discoverer.cellular-automata.com/}.  

\footnotetext{%
One can execute these cellular automata in \textit{Mathematica} replacing \texttt{\textbf{ <Rule>}} with the corresponding rule number

\texttt{%
ListAnimate[ArrayPlot[\#]\&/@CellularAutomaton[\\{\indent\indent<Rule>,2,{1,1}},
{RandomInteger[1,{100,100}],0},100]]
}
}%end footnote

\newpage

\textbf{Rule:} 
354830437430697307314658045280649922899653607237\\
152783088733395073850801752918249535088820853655864680729\\
189540963997737594766246170112169867440686203456\\

\begin{table}[h]
	\centering
	%\captionsetup{justification=centering}
	\begin{tabular}{|c|c|c|}
	\hline
	 & $M_E$ & $M_D$ \\
	\hline
	Chaoticity & 62.11  & 12.06  \\
	\hline
	Decrease   & 4.88   & 78.88  \\
	\hline
	Growth     & 33.01  & 9.06  \\
	\hline
	Stability  & 0 & 0 \\
	\hline
	\end{tabular}
	\caption{Euclidean distance to the feature vector of behavioral metrics of the Game of Life is 9.32. The correlation between measures of static and dynamic behavior in this cellular automaton is -0.34}
	\label{tab:measures}
\end{table}

\begin{table}[h]
\centering
\begin{tabular}{ c }
\includegraphics[scale=0.9]{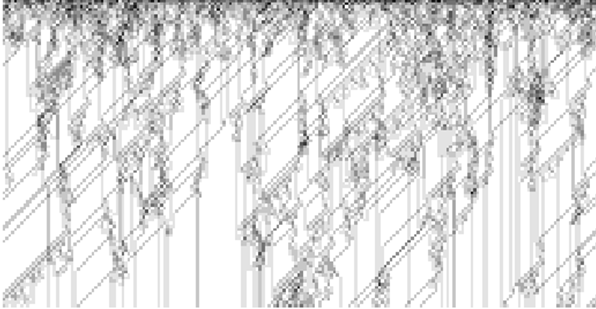}\\
Averaged spacetime evolution\\
\\
\includegraphics[scale=0.7]{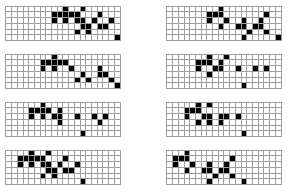} \\
 Identified gliders\\  
\end{tabular}
\end{table}

\newpage

\textbf{Rule:}
196928112803567351078509513317947776313717639009629\\
19233419392303723364585678060118178225231534916460395002491\\
6004629851769274774088586292232688540354568 

\begin{table}[h]
	\centering
	%\captionsetup{justification=centering}
	\begin{tabular}{|c|c|c|}
	\hline
	 & $M_E$ & $M_D$ \\
	\hline
	Chaoticity & 64.45 & 9.28  \\
	\hline 
	Decrease   & 2.54 & 84.80  \\
	\hline
	Growth     & 33.01 & 5.92  \\
	\hline 
	Stability  & 0 & 0 \\
	\hline
	\end{tabular}
	\caption{Euclidean distance to the feature vector of behavioral metrics of the Game of Life is 13.68. The correlation between measures of static and dynamic behavior in this cellular automaton is -0.40}
	\label{tab:measures}
\end{table}

\begin{table}[h]
\centering
\begin{tabular}{ c }
\includegraphics[scale=0.8]{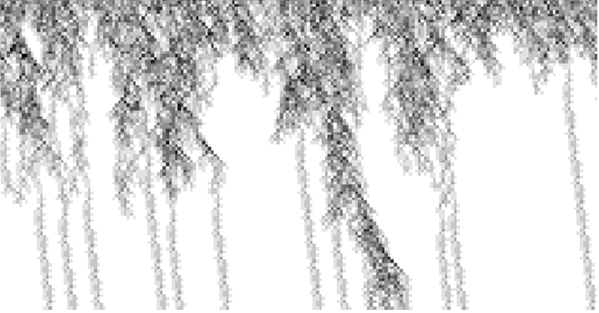}\\
Averaged spacetime evolution\\
\\
\includegraphics[scale=0.8]{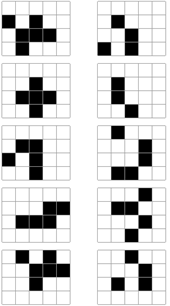} \\
 Identified gliders\\  
\end{tabular}
\end{table}

\newpage

\textbf{Rule:}
25369628583304459989446065099150613536215022807637650\\
13218910019118617632623181726351808015804669971129335990123\\
389394577484439270322287946219773078676008

\begin{table}[h]
	\centering
	%\captionsetup{justification=centering}
	\begin{tabular}{|c|c|c|}
	\hline
	 & $M_E$ & $M_D$ \\
	\hline
	Chaoticity & 65.63 & 5.53  \\
	\hline 
	Decrease   & 3.91 & 90.54  \\
	\hline
	Growth     & 30.47 & 4.00  \\
	\hline 
	Stability  & 0 & 0 \\
	\hline
	\end{tabular}
	\caption{Euclidean distance to the feature vector of behavioral metrics of the Game of Life is 19.13. The correlation between measures of static and dynamic behavior in this cellular automaton is -0.42.  }
	\label{tab:measures}
\end{table}

\begin{table}[h]
\centering
\begin{tabular}{ c }
\includegraphics[scale=0.8]{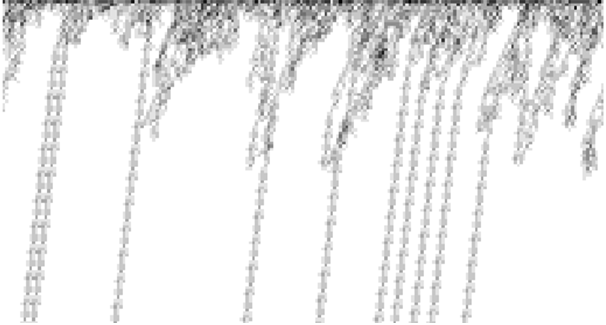}\\
Averaged spacetime evolution\\
\\
\includegraphics[scale=0.8]{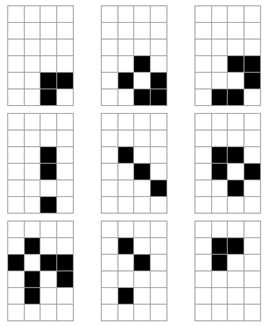} \\
 Identified glider\\  
\end{tabular}
\end{table}

\newpage
\section*{\label{ack}Acknowledgements}
We wish to thank Jan Baetens, Hector Zenil, Alyssa Adams, and Nima Dehghani for their helpful comments. We appreciate the support of the Physics and Mathematics in Biomedicine Consortium. We also wish to thank Todd Rowland for his encouragement and continued interest in the project

% Begin appendix sections

% It is important that
% \textit{all names of journals be spelled out in full}, and italicized.
% (Use \textit{Physical Review Letters}, not \textit{Phys.\ Rev.\ Lett.},
% and \textit{Journal of Computer and System Science}, not
% \textit{J.~Comput.\ Sys.\ Sci}.)

%Give volume numbers in boldface (do not write the word ``volume''
%explicitly). Include the issue number of the volume only if the
%journal is renumbered with each issue.  If the issue number is
%necessary, place it in parentheses immediately following the volume
%number, but not in bold; for example, \textbf{4}(1). Give dates in
%parentheses; include months only when necessary, spelling them
%out in full. Give starting \textit{and ending} page numbers for papers.

%Titles of books (i.e., published material with ISBN numbers)
%should be italicized. Names and cities of publishers and dates
%of publication should always be given. Complete addresses should
%be given for small publishers. Conference proceedings that are
%distributed through ordinary publishers should be referenced like
%books.


\begin{thebibliography}{99}
% The number of 9s indicates how wide the reference
% numbers must hang for all references to be aligned properly.


%To cater to a wide variety of disciplines, it is important that
%\textit{all names of journals be spelled out in full}, and italicized.
%(Use \textit{Physical Review Letters}, not \textit{Phys.\ Rev.\ Lett.},
%and \textit{Journal of Computer and System Science}, not
%\textit{J.~Comput.\ Sys.\ Sci}.)

%Give volume numbers in boldface (do not write the word ``volume''
%explicitly). Include the issue number of the volume only if the
%journal is renumbered with each issue.  If the issue number is
%necessary, place it in parentheses immediately following the volume
%number, but not in bold; for example, \textbf{4}(1).


%First Author and Second Author, ``A Review Article,'' \textit{Full Name
%of Journal}, \textbf{volume} (date) page--page.

% Use \textit{...} for italics, \textbf{...} for boldface.
% Do not explicitly use the word ``volume''
% The full year (e.g., 1991) should be used for the date.
% Use -- to get an appropiate dash between page numbers.


%\textit{Full titles of papers should be given}. They should be enclosed
%in quotation marks, with all important words capitalized. They should
%be followed by a comma inside the quotation marks.

\bibitem{nks}
S. Wolfram, \textit{A New Kind of Science}, Wolfram Media, Champaign, IL, 2002.

\bibitem{stat-mech-ca}
S. Wolfram, "Statistical Mechanics of Cellular Automata," \textit{Reviews of Modern Physics},
\textbf{55}(3), 1983, pp. 601-644.

\bibitem{universality-ca}
S. Wolfram, "Universality and Complexity in Cellular Automata," \textit{Physica D: Nonlinear Phenomena}, \textbf{10}(1), 1984, pp. 1-35. 

\bibitem{collision-computing}
A. Adamatzky and J. Durand-Lose, "Collision Based Computing," section of the \textit{Handbook of Natural Computing}, Springer, 2012, pp. 1949-1978 

\bibitem{mol-assembly-ca}
M. Nilsson and S. Rasmussen, "Cellular Automata for Simulating Molecular Self-Assembly," \textit{Discrete Mathematics and Theoretical Computer Science}, 2003, pp. 31-42.

\bibitem{sierpinski-dna}
P. Rothemund, N. Papadakis and E. Winfree, "Algorithmic Self-Assembly of DNA Sierpinski Triangles," \textit{PLOS Biology}, 2004.

\bibitem{amorphous-comp}
H. Abelson, D. Allen, D. Coore, C. Hanson, E. Rauch, G. J. Sussman, G. Homsy, J. Thomas F. Knight and R. W. Radhika Nagpal, "Amorphous Computing," \textit{Communications of the ACM}, \textbf{43}(5), 2000, pp. 74-82.

\bibitem{dna-self}
M. Hirabayashi, S. Kinoshita, S. Tanaka, H. Honda, H. Kojima and K. Oiwa, "Cellular Automata Analysis on Self-Assembly Properties in DNA Tile Computing," \textit{Lecture Notes in Computer Science}, \textbf{7495}, 2012,  pp. 544-553.

\bibitem{rule-simulation}
M. Hwang, M. Garbey, S. A. Berceli and R. Tran-Son-Tay, "Rule-Based Simulation of Multi-Cellular Biological Systems—A Review of Modeling Techniques," \textit{Cellular and Molecular Bioengineering}, \textbf{2}(3), 2009, pp. 285-294.

\bibitem{bio-modelling}
G. B. Ermentrout and L. Edelstein-Keshet, "Cellular Automata Approaches to Biological Modelling," \textit{Journal of Theoretical Biology}, \textbf{160}, 1993, pp. 97-133.

\bibitem{handbook}
G. Rozenberg , T. Bäck and J. Kok, (editors), \textit{Handbook of Natural Computing}, Springer, 2012, pp. 1-287, pp. 1168.

\bibitem{biochem-ca}
L. B. Kier, D. Bonchev and G. A. Buck, "Modeling Biochemical Networks: A Cellular-Automata Approach," \textit{Chemistry and Biodiversity}, \textbf{2}(2), 2005, pp. 233-243.

\bibitem{game-of-life}
M. Gardner, "Mathematical Games - The Fantastic Combinations of John Conway's New Solitaire Game "Life"," \textit{Scientific American}, \textbf{223}, 1970, pp. 120-123.

\bibitem{lyapunov-1}
J. M. Baetens and B. De Baets, "Towards the Full Lyapunov Spectrum of Cellular Automata," \textit{AIP Conference Proceedings}, \textbf{1389}(1), 2011, pp. 981-986.

\bibitem{lyapunov-2}
J.M. Baetens and J. Gravner, "Stability of cellular automata trajectories revisited: branching walks and Lyapunov profiles," 2014. [Online]. Available: \url{http://arxiv.org/pdf/1406.5553.pdf}.

\bibitem{kolmogorov}
H. Zenil and E. Villareal-Zapata, "Asymptotic Behaviour and Ratios of Complexity in Cellular Automata," \textit{International Journal of Bifurcation and Chaos}, \textbf{23}(9), 2013.

\bibitem{genetic-search}
M. Mitchell, J. P. Crutchfield and P. T. Hraber, "Evolving Cellular Automata to Perform Computations: Mechanisms and Impediments," \textit{Physica D: Nonlinear Phenomena}, 75(1), 1994, pp. 361-391.

\bibitem{intro-to-ga}
M. Mitchell. \textit{An Introduction to Genetic Algorithms}. MIT press, 1998

\bibitem{ca-signal}
J. Wurthner, A. Mukhopadhyay, and Claus-Jürgen Peimann. "A cellular automaton model of cellular signal transduction." Computers in biology and medicine \textbf{30}(1), 2000, pp. 1-21.






%\bibitem{preprint}
%A. Author, ``Title of Preprint,'' Institution name, address (date if appropriate).



%\bibitem{program}
%Name, \texttt{Program Name} [computer program in specified \textsc{language}
%running on particular computer systems] (available from Organization,
%Address).
% Typewriter or small caps font may be used for computer program names.

\end{thebibliography}
\end{document}